\documentstyle[12pt,epsfig,aaspp4]{article}

\received{2003 March 26}
\begin{document}

\newcommand{\lya}{Lyman~$\alpha$}
\newcommand{\lyb}{Lyman~$\beta$}
\newcommand{\za}{$z_{\rm abs}$}
\newcommand{\ze}{$z_{\rm em}$}
\newcommand{\cmtwo}{cm$^{-2}$}
\newcommand{\nhi}{$N$(H$^0$)}
\newcommand{\nzn}{$N$(Zn$^+$)}
\newcommand{\ncr}{$N$(Cr$^+$)}
\newcommand{\degpoint}{\mbox{$^\circ\mskip-7.0mu.\,$}}
\newcommand{\halpha}{\mbox{H$\alpha$}}
\newcommand{\hbeta}{\mbox{H$\beta$}}
\newcommand{\hgamma}{\mbox{H$\gamma$}}
\newcommand{\kms}{\,km~s$^{-1}$}      
\newcommand{\minpoint}{\mbox{$'\mskip-4.7mu.\mskip0.8mu$}}
\newcommand{\mv}{\mbox{$m_{_V}$}}
\newcommand{\Mv}{\mbox{$M_{_V}$}}
\newcommand{\peryr}{\mbox{$\>\rm yr^{-1}$}}
\newcommand{\secpoint}{\mbox{$''\mskip-7.6mu.\,$}}
\newcommand{\sqdeg}{\mbox{${\rm deg}^2$}}
\newcommand{\squig}{\sim\!\!}
\newcommand{\subsun}{\mbox{$_{\twelvesy\odot}$}}
\newcommand{\et}{et al.~}

\def\ltsima{$\; \buildrel < \over \sim \;$}
\def\simlt{\lower.5ex\hbox{\ltsima}}
\def\gtsima{$\; \buildrel > \over \sim \;$}
\def\simgt{\lower.5ex\hbox{\gtsima}}
\def\arcs{$''~$}
\def\arcm{$'~$}
%
%
\title{THE C~IV MASS DENSITY OF THE UNIVERSE AT REDSHIFT 5~\altaffilmark{1}}
\author{\sc Max Pettini}
\affil{Institute of Astronomy, Madingley Road, Cambridge, CB3 0HA, UK}
\author{\sc Piero Madau}
\affil{Department of Astronomy and Astrophysics, University of California, Santa Cruz,
CA 95064}
\author{\sc Michael Bolte and Jason X. Prochaska}
\affil{University of California Observatories/Lick Observatory,
University of California, Santa Cruz, CA 95064}
\author{\sc Sara L. Ellison}
\affil{P. Universidad Cat\'olica de Chile, Casilla 306, Santiago 22, Chile and}
\affil{European Southern Observatory, Casilla 19001, Santiago 19, Chile}
\author{\sc Xiaohui Fan}
\affil{Steward Observatory, The University of Arizona, Tucson, AZ 85721}

\altaffiltext{1}{Based on data obtained at the W. M. Keck Observatory which
is operated as a scientific partnership among the California 
Institute of Technology, the
University of California, and NASA, and was made possible by
the generous financial support of the W.M. Keck Foundation. }

\begin{abstract}
In order to search for metals in the Lyman\,$\alpha$ forest at
redshifts $z_{\rm abs} > 4$, we have obtained spectra of 
high signal-to-noise ratio and
moderately high resolution of three QSOs at $z_{\rm em} > 5.4$
discovered by the Sloan Digital Sky Survey. 
These data allow us to probe to
metal enrichment of the intergalactic medium at early times 
with higher sensitivity than previous studies.
We find 16 C~IV absorption systems with column densities
$\log N{\rm (C~IV)} = 12.50 - 13.98$ over a total redshift path
$\Delta X = 3.29$. In the redshift interval
$z_{\rm} = 4.5 - 5.0$, where our statistics are most reliable,
we deduce a comoving mass density of 
C$^{3+}$ ions $\Omega_{\rm C~IV} = (4.3 \pm 2.5) \times 10^{-8}$
(90\% confidence limits) for absorption systems with 
$\log N{\rm (C~IV)} \geq 13.0$ (for an Einstein-de Sitter cosmology with
$h = 0.65$).
This value of $\Omega_{\rm C~IV}$ is entirely consistent with
those measured at $z < 4$; we confirm the earlier finding by Songaila
(2001) that neither the column density distribution of C~IV absorbers
nor its integral show significant redshift evolution over a period
of time which stretches from $\sim 1.25$ to $\sim 4.5$\,Gyr after the big bang.
This somewhat surprising conclusion may be an indication that 
the intergalactic medium was enriched in metals
at $z \gg 5$, perhaps by the sources responsible for its reionization.
Alternatively, the C~IV systems we see may be associated with outflows
from massive star-forming galaxies at later times, while the truly
intergalactic metals may reside in regions of the Lyman\,$\alpha$
forest of lower density than those probed up to now.
\end{abstract}


\section{INTRODUCTION}
Efforts to probe the universe at ever higher redshifts 
have continued to gather momentum in the last
few years. The Sloan Digital Sky 
Survey (SDSS) has led to the discovery of many quasars
(QSOs) at $z > 5$ (Fan et al. 2003 and references therein)
and searches for normal star forming galaxies at these epochs
have been equally successful, reaching $z \simeq 6.6$ 
(Hu et al. 2002; Yan, Windhorst, \& Cohen 2003; 
Lehnert \& Bremer 2003; Kodaira et al. 2003; 
Stanway, Bunker, \& McMahon 2003). 
One of the main goals of all of these studies is 
to identify the time when the baryons in the 
intergalactic medium (IGM) were reionized by the light of the 
first stars and galaxies. The spectra of SDSS QSOs at $z\simgt 6$ are essentially
black at wavelengths below the \lya\ emission line,
a finding which has been interpreted as the signature of the trailing
edge of the cosmic reionization epoch (e.g. Becker et al. 2001; Fan et al. 2002; 
Cen \& McDonald 2002; White et al. 2003). 
Yet, the recent detection by the {\it Wilkinson 
Microwave Anisotropy Probe} ({\it WMAP}) of a large optical depth to 
Thomson scattering, $\tau_e=0.17\pm 0.04$, suggests that the universe was reionized 
at higher redshifts, $z_{\rm ion}=17\pm 5$ (Kogut et al.  2003; Spergel et al. 2003).
If confirmed, this would be an indication of significant star-formation
activity at very early times.

Beyond $z \simeq 4$, the information provided by 
the \lya\ forest itself becomes progressively more difficult to 
interpret because of the severe line 
blending and rapidly increasing optical depth which 
leave little signal in QSO spectra. 
In order to follow the evolution of the 
IGM to earlier times we have to rely on the metal lines
associated with the \lya\ forest, the most common of which
is the C~IV~$\lambda\lambda 1548.2041,1550.7812$ doublet.
The standard for this work has been set by the analysis
of Songaila (2001). By bringing together measurements from
the spectra of 32 QSOs, Songaila was able
to follow the evolution of the column density distribution
of C~IV absorbers, $f(N)$,  over the redshift interval 
$1.5 \leq z \leq 5$. In today's `consensus' cosmology,
$\Omega_{\rm M}=0.3$, $\Omega_\Lambda=0.7$, 
$H_0 = 65$\,km~s$^{-1}$~Mpc$^{-1}$, 
this redshift range
corresponds to a time interval from 4.5 to 1.25\,Gyr after
the big bang. The surprising result is that
no evolution can be discerned in $f(N)$, nor in its integral
which gives the mass density of C$^{3+}$ ions, 
$\Omega_{\rm CIV}$ (expressed as a fraction of the 
critical density). 
Taken at face value, this finding may suggest
that most of the IGM metals were already in place at the highest
currently observable redshifts, and may thus point to an {\it early} enrichment
epoch by outflows from low-mass subgalactic systems (Madau, Ferrara, \& Rees 2001). 
Alternatively, the density of C$^{3+}$ ions may not reflect in a simple way the overall 
density of metals in the IGM if, for example, {\it late} winds from massive
Lyman break galaxies are the source of the strongest C~IV absorption systems
(Haehnelt 1998; Adelberger et al. 2003).
It then becomes extremely important to push the study of the C~IV `forest' 
to higher redshifts not only to distinguish between these different chemical 
evolution scenarios, but also to understand the mechanisms by which 
metals are distributed from their stellar birthplaces and mixed within the IGM.  

The sample of QSOs analyzed by Songaila (2001) 
was assembled as the first results from the SDSS were 
beginning to appear in the public domain,
and therefore included relatively few objects at $z > 5$. 
Consequently, her statistics
on $f(N)$ and $\Omega_{\rm CIV}$ are least secure at the
highest redshifts. In this paper, we add to Songaila's
work with deep observations of three recently discovered
SDSS QSOs, all at $z > 5$, obtained with the aim of improving the 
statistics of the column density distribution by:
(a) increasing the sample, (b) reaching to lower 
values $N$(C~IV), and (c) considering the likely
corrections due to sample incompleteness.
Our main conclusion is that we confirm the previously
reported lack of evolution in $\Omega_{\rm CIV}$.
The observations are described in \S2, while in \S3 we provide
measurements of the C~IV absorbers detected in the
three QSO spectra. In \S4 and \S5 we analyze our sample
and compare our findings with those reported by Songaila
(2001). Finally, in \S6 we briefly discuss possible 
interpretations of these results and their implications 
for the origin of C~IV absorption at high redshifts.\\

\section{OBSERVATIONS AND DATA REDUCTION}

The spectra of the three QSOs were recorded 
with the Echelle Spectrograph and Imager
(ESI; Sheinis et al. 2000) on the Keck~II telescope
in January and February of 2002; relevant details
of the observations are collected in Table 1.
With its combination of high efficiency at red and near-IR wavelengths,
wide wavelength coverage (from 4000\,\AA\ to 1\,$\mu$m),
and moderately high resolution ($R \simeq 6500$) 
ESI is well suited to the aims of the present work.

The QSOs SDSS\,0231$-$0728 (Anderson et al. 2001) at
$z_{\rm em} = 5.421$\footnote{The values of emission redshift 
quoted in this paper were measured from the onset of the 
\lya\ forest in the ESI data presented here and therefore 
differ slightly from those given in the original discovery
papers which were based on lower resolution spectra.}
and SDSS\,0836+0054 (Fan et al. 2001b) at $z_{\rm em} = 5.803$
were observed with a 0.75\,\arcs wide entrance slit, which
projects to approximately four $11.5$\,km~s$^{-1}$ pixels in
the dispersion direction. For SDSS\,1030+0524 (Fan et al. 2001b)
at $z_{\rm em} = 6.305$, a 1\,\arcs wide entrance slit
was employed. For all observations the slit was aligned at the
parallactic angle, and the airmass was always less than 1.5\,.
Reference spectra of internal lamps
were used for wavelength calibration and flat-fielding. 
Observations of the smooth spectrum white dwarf star G191B2B,
obtained on each night, provided a template 
for dividing out the numerous telluric absorption lines 
which mar ground-based spectra in the far red and near-IR;
they were also used to place 
the QSO spectra on an absolute flux scale
(Massey et al 1988; Massey \& Gronwall 1990).

The individual two-dimensional ESI images (recorded
with exposure times of either 1200, 1800, or 2700\,s)
were processed using custom IDL routines written
by one of us with the specific aim of maximising
the accuracy of background subtraction\footnote{The 
ESI data processing package is publicly available at\\
http://www2.keck.hawaii.edu/realpublic/inst/esi/ESIRedux/index.html}; 
this is often the limiting factor in deep spectroscopy
of faint sources at long optical wavelengths, where
line emission from the night sky dominates the signal.

After flat-fielding, wavelength calibration and
background subtraction, the individual one-dimensional
spectra were mapped onto a common, vacuum heliocentric, 
wavelength scale before being added together to produce
the final spectra, shown in Figure 1.  

Corresponding variance spectra
are also produced by the data reduction software.
In column (6) of Table 1 we list typical values of the 
signal-to-noise ratio (S/N) in the QSO continuum in
the wavelength regions where we searched for C~IV doublets 
(see \S3). Generally, the S/N is highest near 
the \lya\ emission line and decreases at longer 
wavelengths. In the two best observed QSOs,
SDSS\,0836+0054 and SDSS\,1030+0524, the 
long exposure times (see column (5) of Table 1)
resulted in S/N\,$\simgt 60$ and 40 respectively
at wavelengths between \lya\ emission and 9000\,\AA.
The resolution of the spectra, as measured from the 
widths of night sky emission lines, is
$\sim 1.3$\,\AA~FWHM ($\simeq 45$\,km~s$^{-1}$), sampled
with $\sim 4$ wavelength bins for the QSOs
SDSS\,0231$-$0728 and SDSS\,0836+0054
(and $\sim 30$\% coarser for SDSS\,1030+0524).
With their combination of
high S/N and resolution, the spectra used in this study
are some of the best published of very high 
redshift QSOs, as can be appreciated from Figure 1 and from 
the last column of Table 1 which lists the
$3 \sigma$ detection limits for the rest frame
equivalent widths of unresolved C~IV lines.

The final steps in the data reduction involved
correcting for atmospheric absorption by diving
the QSO spectra by that of the smooth spectrum 
star (suitably normalized), and fitting the 
QSO continuum. Both steps were carried out using
the Starlink software package DIPSO\footnote{ 
See http://www.starlink.rl.ac.uk/star/docs/sun50.htx/sun50.html}.
The end result are normalised
QSO spectra which could then be searched for C~IV
absorption.\\

\section{C~IV ABSORPTION LINES AT HIGH REDSHIFT}

The spectra of the three QSOs were visually inspected
independently by two of us for pairs of absorption lines
with the correct separation and relative strengths to be C~IV doublets.
After various trials (including the simulations described in \S5 below),
we decided to restrict the search to two wavelength regions.
The first region extends from the \lya\ emission line of the QSO
longward to 8940\,\AA, near the onset of the atmospheric A-band.
The second region is a relatively small gap, between 9200 and 9300\,\AA,
which is free of strong atmospheric absorption and sky emission lines
(and, for this reason, is being exploited in narrow-band searches
for high redshift \lya\ emitters---see, for example, Hu et al. 2002).
The reason for restricting ourselves to these wavelength intervals
is that at other wavelengths strong atmospheric absorption reduces
significantly the S/N achieved. Even with 
the best efforts to divide out the atmospheric lines, our
sensitivity to C~IV absorption is much reduced here;  simulations
confirmed that we could only recover the strongest C~IV doublets, and
with only partial success.
Although this choice effectively imposes a limit of $z_{\rm abs} \leq 5.0$
to the redshift range over which we can study C~IV, we nevertheless
prefer to concentrate our analysis on regions where our detection
limit is relatively uniform.
The simulations described in \S5 show that in these regions we are
essentially complete for C~IV absorption lines with rest frame
equivalent width $W_0 \geq 40$\,m\AA.

Once C~IV doublets had been identified, 
their normalized line profiles were fitted
with theoretical Voigt profiles 
using the VPFIT package.\footnote{VPFIT
is available at http://www.ast.cam.ac.uk/\~\,rfc/vpfit.html.}
VPFIT deconvolves the composite absorption profiles
into the minimum number of discrete components
and returns for each the most likely values of
redshift \za, Doppler width $b$ (km~s$^{-1}$),
and column density $N$(C~IV) (cm$^{-2}$)
by minimizing the difference
between observed and computed profiles.
The profile decomposition takes into
account the instrumental
point spread function of ESI.
Vacuum rest wavelengths and $f$-values of the
C~IV transitions are from the recent compilation
by Morton (2003).
We now briefly discuss each QSO in turn.

\subsection{SDSS\,0231$-$0728}

The peak of the \lya\ emission line in this QSO,
at 7806\,\AA, corresponds to \za=4.0420 for
C~IV$\lambda 1548.2041$; thus, we can search for
C~IV doublets in the redshift intervals 
$4.0420 \leq z_{\rm abs} \leq 4.7745$ (7806--8940\,\AA)
and $4.9424 \leq z_{\rm abs} \leq 5.0070$ (9200--9300\,\AA).
We find eight C~IV systems, listed in Table 2 and 
reproduced, together with their profile fits, in Figure 2.
The two systems labelled `Marginal' in Table 2 
are cases where we do not feel confident of the identifications
because the weaker member of the doublet is affected by
residuals in the sky subtraction.

\subsection{SDSS\,0836+0054}
The higher redshift of this QSO (the peak of the
\lya\ emission line is at 8270\,\AA) means that we
can only search for C~IV doublets over a more restricted
redshift range, from $z_{\rm abs} = 4.3417$ to 4.7745
(8270--8940\,\AA), as well as 
$4.9424 \leq z_{\rm abs} \leq 5.0070$ (9200--9300\,\AA).
We find seven C~IV absorption systems, one of which
consists of (at least) three separate components
(see Table 3 and Figure 3).

\subsection{SDSS\,1030+0524}
This is the highest redshift QSO among the three studied here;
with the peak of the \lya\ emission line occurring 
at 8880\,\AA, we have 
only 60\,\AA\ of clear continuum before the onset of
the atmospheric A-band. We find no C~IV systems in this
narrow redshift interval (\za = 4.7357 -- 4.7745), although
one is detected between 9200 and 9300\,\AA, at
\za = 4.94866 (see Table 4 and Figure 4).  

The errors quoted in Tables 2, 3, and 4 are 
the $1 \sigma$ estimates returned by VPFIT
on the basis of the error spectra provided
to the fitting program. These spectra are shown 
as a line near the zero level in each panel of
Figures 2, 3, and 4, and can be seen to be a reasonable
representation of the rms deviations from the
continuum level away from strong sky lines
(the residuals from the subtraction of the
sky lines can sometimes amount to many times the estimated
value of sigma, reflecting systematic, rather than random,
errors in the sky subtraction).
However, VPFIT does not take into account the possibility
that what we see as an individual C~IV line
is actually an unrecognized blend of more than one 
absorption component. This is a common problem
in the analysis of interstellar (or in this case intergalactic)
absorption lines,
made worse here by the relatively coarse resolution
of ESI. Specifically, the ESI point spread function
of FWHM\,=\,45\,km~s$^{-1}$ corresponds to an instrumental
Doppler parameter $b_{\rm instr} = 27$\,km~s$^{-1}$.
This is significantly larger than the typical $b$-values
of C~IV lines at lower redshifts ($z = 2 - 3$); for example
Rauch et al. (1996) reported 
$b_{\rm median}$(C~IV)\,$= 10.6$\,km~s$^{-1}$,
while Ellison et al. (2000) found 
$b_{\rm median}$(C~IV)\,$= 13$\,km~s$^{-1}$, both
from higher resolution ($5 - 8$\kms\ FWHM)
spectra than those considered here. 

There are several hints in our data that we may
underestimating the complexity of the C~IV
absorption lines. 
First, the values of $b$ we deduce are all
larger than the median values at $z = 2-3$,
as referenced above. We consider it much more likely 
that this is an effect due to the limited spectral
resolution of our data, rather than a genuine redshift evolution
in the intrinsic line widths.
Second, 
unrecognized velocity structure
may be the reason why the fits to the some
of our C~IV systems are poor (examples are the 
\za\ = 4.685 complex and the \za\ = 4.99695
system in Figure 3). 
In general, the result of blending several components
into one unresolved feature is 
an {\it under}estimate of the column density,
because narrow saturated components, if they exist, could
easily be masked by broader ones (Nachman \& Hobbs 1973).
The \za\ = 4.66874 system in Figure 3
is an example where the C~IV doublet ratio is indicative of 
saturation. VPFIT converges to a value of $b_{\rm CIV}$ very much
less than $b_{\rm instr}$; therefore, in this case
$b_{\rm CIV}$ remains undetermined and we quote in Table 3
a lower limit to $N$(C~IV) based on the equivalent width
of $\lambda 1550.7812$, the weaker member of the doublet,
assuming no saturation.
In conclusion, given the limited resolution of ESI, 
the values of $N$(C~IV) derived here
should strictly be considered as lower limits, 
a point which we shall keep
in mind in the interpretation of the results.\\

\section{THE MASS DENSITY OF C~IV}

The mass density of C~IV ions, expressed as a fraction of
the critical density today, $\rho_{\rm crit} = 1.89 
\times 10^{-29}\,h^2$\,g~cm$^{-3}$,
can be calculated as (e.g. Lanzetta 1993)
\begin{equation}
\Omega_{\rm CIV} = 
\frac{H_0 \, m_{\rm CIV}}{c \, \rho_{\rm crit}}\int N f(N) {\rm d}N, 
\end{equation}
where $H_0 = 100\,h$\kms~Mpc$^{-1}$ is the Hubble constant,
$m_{\rm CIV}$ is the mass of a C~IV ion, $c$ is the speed of light,
and $f(N)$ is the number of C~IV absorbers per unit column density per unit 
absorption distance $X(z)$. This last quantity is used to remove 
the redshift dependence in the sample: absorbers with constant comoving space 
density and constant proper size have a constant number density per unit 
absorption distance along a line of sight. In a flat Friedmann universe with matter
and vacuum density parameters today $\Omega_M$ and $\Omega_\Lambda$, the absorption 
distance is given by

\begin{equation}
X(z)=\int_0^z dz'\,{(1+z')^2\over
[\Omega_M(1+z')^3+\Omega_\Lambda]^{1/2}}=
{2\over 3\Omega_M}\,\{[\Omega_M(1+z)^3+\Omega_\Lambda]^{1/2}-1\}.
\end{equation}

In the case of an Einstein-de Sitter universe with ($\Omega_M, \Omega_\Lambda)=(1, 0)$, 
we recover the standard expression $X(z)=2[(1+z)^{3/2}-1]/3$ (Tytler 1987).

Our total sample consists of 16 or 12 C~IV systems,
depending on whether we include the marginal detections
or not. Although these absorbers span a factor of $\sim 30$
in column density, from $\log N$(C~IV) = 12.50 to 13.98
(see Tables 2, 3 and 4), our completeness
drops to below 50\% for $\log N$(C~IV)\,$\simlt 13.0$ (\S5).
It is not possible to determine $f(N)$ on the
basis of such limited statistics.
However, we can still deduce an estimate of 
$\Omega_{\rm CIV}$ using the approximation
\begin{equation}
\int N f(N) {\rm d}N = 
\frac{\sum_{i} N_{i} {\rm (C~IV)}}{\Delta X}
\end{equation}
(e.g. Storrie-Lombardi, McMahon, \& Irwin 1996),
where $\Delta X$ is the total absorption distance covered by our spectra.
In the following we will adopt an Einstein-de Sitter cosmology for comparison
with earlier work; then $\Delta X$ is given by
\begin{equation}
\Delta X = 
\frac{2}{3}~
\sum_{j} [(1+z_j^{\rm max})^{3/2} - (1+z_j^{\rm min})^{3/2}].
\end{equation}
The summation in equation (4) is over 
the $j$ redshift intervals $\Delta z$ where we can 
detect C~IV systems in our data; 
as explained in \S3, we have two such intervals in each 
of the three sight-lines to the QSOs in Table 1.
With $h = 0.65$, equations (1) and (3) then lead to:
\begin{equation}
\Omega_{\rm CIV} = 1.75 \times 10^{-22}~ 
\frac{\sum_{i} N_{i}{\rm (C~IV)}}{\Delta X}.
\end{equation}
For the error in $\Omega_{\rm CIV}$, $\delta\Omega_{\rm CIV}$,
we adopt the estimator proposed by Storrie-Lombardi et al. (1996):
\begin{equation}
(\delta\Omega_{\rm CIV})^2 = 
\frac{\sum_{i} [N_{i}{\rm (C~IV)}]^2}
{\left [ \sum_{i} N_{i} {\rm (C~IV)} \right ] ^2}.
\end{equation}
For consistency with Songaila (2001), in what follows
we quote errors which correspond to formal 90\%
confidence limits, that is $\delta\Omega_{\rm CIV} \times 1.64$
(assuming a gaussian distribution).
We also estimated errors using the bootstrap method
(Efron \& Tibshirani 1993) and found them to be about
65\% of the values obtained from eq.(6). We adopt the 
values from eq.(6), as they are likely to be more realistic.

The results are collected in Table 5.  
Over the redshift interval $z = 4.0 - 5.0$ covered
by our spectra, the total redshift path is
$\Delta X = 3.29$. Summing the column densities of all
16 absorption systems detected, we obtain 
$\Omega_{\rm CIV} = (2.4 \pm 1.2) \times 10^{-8}$
at a mean $\langle z \rangle = 4.568$.
Excluding the four `marginal' systems in Tables 2, 3, and 4
would have a very small effect on 
$\sum_{i} N_{i} {\rm (C~IV)}$, and therefore
$\Omega_{\rm CIV}$, because relatively low column densities 
are associated with these uncertain detections.
On the other hand, $\Omega_{\rm CIV}$ may have been
underestimated if some of the C~IV doublet lines
include strongly saturated components, as discussed
above.

In her analysis, Songaila (2001) divided her sample
in bins of $\Delta z = 0.5$. The only such bin where
we have sufficient statistics to compare with her
results is between $z = 4.5$ and 5.0, which includes 11
out of the 16 C~IV systems detected here. For this
subsample of our data, $\Delta X = 1.87$, 
$\langle z \rangle = 4.688$
and $\Omega_{\rm CIV} = (3.6 \pm 2.1) \times 10^{-8}$.
For comparison, Songaila's (2001) survey in this
redshift bin covered $\Delta X = 5.36$ at a very similar
mean redshift $\langle z \rangle = 4.655$,
and yielded 
$\Omega_{\rm CIV} = (2.5^{+1.9}_{-1.4}) \times 10^{-8}$
(see Table 5 and Figure 5).
Once corrected for incompleteness, we will see below that our estimate 
of $\Omega_{\rm CIV}$ is $\sim 75$\% higher than 
Songaila's earlier value, although the two measurements 
are mutually consistent within the errors. 

Our density of absorbers per unit redshift path
is two times higher than that of Songaila (2001),
presumably reflecting the fact that the high S/N ratios of our spectra
allow us to reach further down the column density 
distribution of C~IV. 
Songaila's best fit to the $f(N)$ distribution in the redshift range $2.90 < z <3.54$ 
is $f(N)=10^{-12.4} N_{13}^{-1.8}$, where $N_{13}$(C~IV) is measured in units of 
$10^{13}\,$cm$^{-2}$. 
If this relatively steep slope of $f(N)$ 
persists to the higher redshifts probed here,
by integrating over the column density
we would expect 
$f(>N)=5\,N_{13}^{-0.8}$, 
that is about 5 absorbers with $\log N \geq 13$ 
per unit absorption distance. 
Over the redshift interval $z=4.0-5.0$ covered by our spectra, 
we have identified between 4 and 4.6 systems 
(depending on whether marginal detections are included or not)
with $\log N \geq 13$ per unit $\Delta X$,
in very good agreement with the expected value.
Thus, within the limited statistics of
our sample, it appears that there is little evolution in the density of C~IV
systems with $\log N {\rm (C~IV)} \geq 13.0$. \\

\section{TESTS FOR COMPLETENESS OF C~IV DETECTIONS}

It is evident from the results in Tables 2, 3, and 4, 
that our ability to detect C~IV absorption systems 
decreases as $\log N$(C~IV) decreases. Most of
our systems have $\log N$(C~IV)\,=\,13.0--14.0
and only three have $\log N$(C~IV)\,$ < 13.0$,
whereas $f(N)$ still rises at these column densities,
at least at $z \simeq 3$ (Ellison et al. 2000).
In order to assess the completeness of our sample,
we have performed some checks, as follows.

One of us generated a number of fake C~IV systems
comparable to the number detected in 
our spectra of SDSS\,0231$-$0728 and
SDSS\,0836$+$0054 (approximately nine systems
per simulation). The $b$-values of the fake C~IV
lines were drawn at random from the observed
distribution of $b$-values in our data;
similarly their redshifts
were chosen randomly from the redshift ranges
where we could search for C~IV systems.
These fake C~IV doublets were then added to the
real spectra and searched for by one of the two members
of our team who had previously identified
the real C~IV lines.
Given the visual, rather than automatic, character
of our searches, only a limited number of such trials
could be performed. Specifically, we performed three
such trials for each of the following values of column density:
$\log N$(C~IV)\,=\, 13.3, 13.5, 13.7, and 14.0 and
six trials for $\log N$(C~IV)\,=\, 13.0\,.
Thus, the results of these tests are only indicative;
they are collected in Table 6.

We found that we could recover essentially all
C~IV doublets so long as $\log N$(C~IV)\,$ \geq 13.3$.
Below this value, however, we quickly become incomplete
because the equivalent width of the weaker member
of the doublet is comparable to residuals from the
subtraction of sky emission lines and from the 
correction for atmospheric absorption, and the chances
of blending with these residuals are high.
Specifically, we were only able to recover
21 out of 47 fake C~IV doublets with 
$\log N$(C~IV)\,=\,13.0. Assuming that this
completeness fraction of 45\% applies to all
systems with $13.0 < \log N$(C~IV)$ < 13.3$ 
in our data\footnote{The correction factors we estimate
here are specific to our data and to our analysis,
and should not be applied to other observations.},
the resulting correction factor to $\Omega_{\rm CIV}$
is +22\%. 
The corresponding correction for 
systems with $\log N$(C~IV)$ < 13.0$ is more
difficult to estimate because our
incompleteness becomes severe at these low
column densities. If we are 10\% complete
in the range $12.7 < \log N$(C~IV)$ < 13.0$,
the corresponding correction to $\Omega_{\rm CIV}$
from this interval alone is +28\%; taking both
corrections into account would raise the value of
$\Omega_{\rm CIV}$ by 50\%.
More generally, if the steep slope of $f(N)$
found by Songaila (2001) continued 
below $\log N$(C~IV)$ = 13.0$, we would expect
roughly comparable contributions
to $\Omega_{\rm CIV}$ from each decade in 
column density.

Songaila (2001) estimated that at redshifts
$z < 4$ incompleteness effects become
significant in her sample only for column densities
$\log N$(C~IV)$ < 13.0$. The simulations 
described above then suggest
that the values of $\Omega_{\rm CIV}$ from our study
should be multiplied by a factor of 1.22 for a meaningful
comparison with the values at $z < 4$. The resulting
$\Omega_{\rm CIV} = (4.3 \pm 2.5) \times 10^{-8}$ derived here
for the interval $z = 4.5-5.0$ (where the error
does {\it not} take into account the uncertainty in the 
incompleteness correction) can then be seen from Figure 5
to be entirely consistent with the values measured at lower 
redshifts. Quantitatively, the mean 
of the five values of $\Omega_{\rm CIV}$ at $z < 4$ from the 
survey by Songaila (2001)
is $\langle \Omega_{\rm CIV}\rangle_{1.5 < z < 4} = 5.3  \times 10^{-8}$, 
with a standard deviation $\sigma = 1.4 \times 10^{-8}$;
the value deduced here for $z = 4.5-5.0$ is within 
$1 \sigma$ of $\langle \Omega_{\rm CIV}\rangle_{1.5 < z < 4}$.

\section{DISCUSSION}

The principal conclusion from this work is that
we confirm the findings by Songaila (2001) that
the column density distribution
$f(N)$ and its integral $\Omega_{\rm C~IV}$,
which measures the mass density of C~IV ions in the intergalactic medium, 
evidently remain approximately constant over an interval of time which
stretches from $\sim 1.25$ to $\sim 4.5$\,Gyr
after the big bang ($z = 5 - 1.5$). 

A straightforward interpretation of these findings is that most of the IGM metals 
were already in place at the highest currently observable redshifts; it seems unlikely
that an invariant distribution could be caused by compensating variations in the 
metallicity and ionization parameter. 
In currently popular hierarchical clustering scenarios for the formation of 
cosmic structures, the assembly of galaxies is a bottom--up process in which 
large systems result from the merging of smaller subunits. In these theories 
subgalactic halos with masses comparable to those of present--day dwarf ellipticals
form in large numbers at very early times. Their gas condensed rapidly due to
atomic line cooling, and became self-gravitating: massive stars formed with 
some (perhaps top-heavy) initial mass function, synthesized heavy-elements, and
exploded as supernovae (SNe) after a few $\times 10^7\,$yr, enriching the
surrounding medium. It is a simple expectation of the above scenario that the 
energy deposition by SNe in shallow potential wells will disrupt the newly 
formed protogalaxies and blow away metal-enriched baryons from the 
host, causing the pollution of the IGM at early times (e.g. Tegmark, Silk, \& 
Evrard 1993; Gnedin \& Ostriker 1997; Madau, Ferrara, \& Rees 2001; Mori, Ferrara,
\& Madau 2002). These subgalactic stellar systems, possibly aided by a population 
of accreting black holes in their nuclei and/or by an earlier generation of stars 
in even smaller halos (`minihalos' with virial temperatures of only a few hundred 
kelvins, where collisional excitation of molecular hydrogen is the main coolant),
are believed to have generated the ultraviolet radiation and mechanical energy 
that reheated and reionized the universe (e.g. Haiman \& Holder 2003; Loeb \& Barkana 
2001).

An alternative picture involves later enrichment from Lyman break galaxies 
(LBGs) instead.  In a study which combined QSO absorption line spectroscopy
with deep galaxy imaging and spectroscopy in the same fields, 
Adelberger et al. (2003) have shown that the \lya\ forest and LBGs are 
more closely related than
had been suspected previously. Of particular relevance
to the present discussion is the spatial association
of strong C~IV systems with galaxies: essentially all
of the systems with $\log N$(C~IV)$\geq 14$ in the Adelberger 
et al. (2003) sample are found
within $\Delta z = 600$\,\kms\ and 
$\Delta r \sim 200 h^{-1}$\,kpc (proper distance)
of a LBG. Adelberger et al. (2003)
show that these dimensions, both in space and velocity,
are characteristic of the galactic-scale
outflows driven by the star formation activity in LBGs. Although 
the correlation weakens as one moves to
lower column densities of C~IV, it remains significant 
over the full range of values of $N$(C~IV) sampled here
($\log N{\rm (C~IV)} \simgt 13$).

And yet the apparent lack of evolution in $\Omega_{\rm CIV}$
in both scenarios is somewhat puzzling. If 
these metals are truly intergalactic 
and due to early pollution, then one might expect the fraction
of C which is triply ionized to change between $z =5$ and 1.5,
since the physical conditions in the IGM are thought to
have evolved between these epochs. Its large scale structure 
developed dramatically, so that a given optical depth
in the \lya\ forest generally refers to condensations of lower overdensity 
(relative to the mean) at $z = 5$ than at $z = 1.5$ (e.g. Cen 2003).
Perhaps most importantly, the ionizing background
may have changed in both intensity and shape, as the 
comoving density of bright QSOs grew to a peak near 
$z = 2.5$ (Fan et al. 2001a) and if the universe
became transparent at wavelengths below
228\,\AA\ following the reionization of helium at $z \sim 3$
(Bernardi et al. 2003; Vladilo et al. 2003). 
Photons with $\lambda < 228$\,\AA\
have sufficient energy to ionize
C$^{3+}$ and thereby reduce the C~IV/C$_{\rm TOT}$
ratio (Dav\'{e} et al. 1998).

If, on the other hand, the metals are ejected from star-forming LBGs, 
the abundance and ionization fraction of C atoms are likely to depend 
more closely on local conditions, rather than those of the IGM at large. 
As suggested by Adelberger et al. (2003), the approximately constant 
value of $\Omega_{\rm CIV}$ may then simply mirror the behaviour of the 
cosmic star formation rate density, which remains
essentially flat over the redshift interval $z \simeq 1.5 - 4$ (Steidel et al. 1999). 
In this picture, while the mean metallicity of universe grows with cosmic time,
one has to assume that this growth is not reflected in the quantity
$\Omega_{\rm C IV}$, perhaps because the systems which make the larger
contribution to this integral (in current data sets) are associated
with outflowing interstellar gas.
The true intergalactic metals may be those at the low
column density end of $f(N)$, below $\log N$(C~IV)$ \simeq 13.0$
(Haehnelt 1998),
where data are still limited to a few sight-lines to
the brightest QSOs (Ellison et al. 2000).
Note also that the results of Adelberger et al. apply
to galaxies and the IGM at $z = 3$, and it remains to
be established whether a similar picture holds at higher
and lower redshifts.  At $z < 0.9$ Chen, Lanzetta, \& Webb
(2001) do find that C~IV absorption systems are 
clustered around galaxies on velocity scales
of up to $\sim250$\,km~s$^{-1}$ and linear scales
of up to $\sim 100\,h^{-1}$\,kpc, but those systems are
generally stronger than the ones considered here.

It is possible that both enrichment mechanisms are at work, and it is 
also perhaps conceivable that, by coincidence, all complicating effects described 
above might work in opposite directions and compensate each other to 
maintain the approximately invariant $\Omega_{\rm CIV}$
found by Songaila (2001) and confirmed here. These possibilities
can only be assessed quantitatively with detailed
calculations which are beyond the scope of this paper.
From an observational point of view,
improving the sensitivity of the spectroscopy presented in this paper
to include weaker C~IV systems 
is a very challenging task at $z > 4$ , even with 8-10\,m telescopes.
On the other hand, there is an incentive to extend this
work to even higher redshifts. 
Detecting even only the strongest C~IV absorbers 
at $z > 5$ (which will require paying special attention to the
problem of correcting for atmospheric absorption),
would still provide an extremely important probe of the
star formation activity at very early epochs. \\

Support for this work was provided by NSF grant AST-0205738 and
by NASA grant NAG5-11513 (P.M.). 
We are grateful to the ESI team for developing
the efficient instrument which made this study possible;
to the staff of the Keck Observatory for their competent
assistance with the observations; and to Bob Carswell
for providing the VPFIT software. 
We are indebted to Robert Becker, Richard White,
and Michael Strauss who kindly contributed 
most of the ESI exposures of SDSS~1030+0524.
We acknowledge valuable discussions with Martin Haehnelt.
Max Pettini thanks the Instituto de Astrof\'{\i}sica de Canarias 
for their hospitality during a visit when this work was completed.
Finally, we wish to extend special thanks
to those of Hawaiian ancestry on whose sacred mountain 
we are privileged to be guests. Without their generous
hospitality, the observations presented herein would not
have been possible.\\


\clearpage

%
%


\begin{deluxetable}{clllllc}
\tighten
\tablecaption{\textsc{Journal of Observations}}
\tablehead{
  \multicolumn{1}{c}{QSO}
& \multicolumn{1}{c}{$z_{\rm em}$\tablenotemark{a}}
& \multicolumn{1}{c}{$z^{\ast}$\tablenotemark{b}}
& \multicolumn{1}{c}{Date}
& \multicolumn{1}{c}{Integration} 
& \multicolumn{1}{c}{S/N\tablenotemark{c}} 
& \multicolumn{1}{c}{$W_0$($3\sigma$)\tablenotemark{d}} \\
  \colhead{ }
& \colhead{ } 
& \multicolumn{1}{c}{(mag)}
& \colhead{ } 
& \multicolumn{1}{c}{(s)}
& \colhead{ } 
& \multicolumn{1}{c}{(m\AA)}
}

\startdata
SDSS\,0231$-$0728\tablenotemark{e} & 5.421  & 19.19  & 2002--Jan--13 \&  & ~~~10\,800 & 50--20 & 14--35 \\ 
                &        &         & 2002--Feb--07  &            &        &    \\
SDSS\,0836+0054\tablenotemark{f}   & 5.803  & 18.74  & 2002--Jan--13 \&  & ~~~32\,400 & 80--40 & 8--17 \\
                &        &         & 2002--Feb--07  &            &        &    \\
SDSS\,1030+0524\tablenotemark{f}   & 6.305  & 20.05  & 2002--Jan--10,11,12 & ~~~28\,500 & 60--15 & 11--45 \\
\enddata
\tablenotetext{a}{Vacuum heliocentric. Measured from our ESI spectra, 
based on the onset of the \lya\ forest.}
\tablenotetext{b}{Magnitude in the Gunn $z^{\ast}$ filter.}
\tablenotetext{c}{Typical signal-to-noise ratio (per pixel) in the continuum over
the wavelength range of interest.}
\tablenotetext{d}{Corresponding $3 \sigma$ limits for the rest frame 
equivalent width of an unresolved absorption line.}
\tablenotetext{e}{Discovery spectrum reported by Anderson et al. (2001)}
\tablenotetext{f}{Discovery spectrum reported by Fan et al. (2001b).}

\end{deluxetable}


\begin{deluxetable}{cllll}
\tighten
\tablecaption{\textsc{C~IV Absorption Systems in SDSS\,0231$-$0728}}
\tablehead{
  \multicolumn{1}{c}{Number}
& \multicolumn{1}{c}{$z_{\rm abs}$\tablenotemark{a}}
& \multicolumn{1}{c}{$b$}
& \multicolumn{1}{c}{$\log N{\rm (C~IV)}$\tablenotemark{b}}
& \multicolumn{1}{c}{~~Comments} \\
  \colhead{ }
& \colhead{ } 
& \multicolumn{1}{c}{(km~s$^{-1}$)}
& \colhead{ } 
& \colhead{ } 
}

\startdata
1       & $4.1242 \pm 0.0001$    & $57 \pm 13$ & $13.03 \pm 0.05$  & \\ 
2       & $4.13921 \pm 0.00002$  & $13 \pm 4$  & $13.19 \pm 0.02$  & \\
3       & $4.22641 \pm 0.00009$  & $40 \pm 9$  & $13.16 \pm 0.06$  & \\
4       & $4.2991 \pm 0.0001$    & $45 \pm 12$ & $13.34 \pm 0.07$  & \\
5       & $4.5070 \pm 0.0001$    & \ldots      & $12.75:$          & ~~Marginal \\
6       & $4.54420 \pm 0.00008$  & $35 \pm 8$  & $13.18 \pm 0.05$  & \\
7       & $4.5701 \pm 0.0001$    & $58 \pm 13$ & $13.38 \pm 0.06$  & \\
8       & $4.7556 \pm 0.0001$    & $53 \pm 12$ & $13.35 \pm 0.07$  & ~~Marginal \\
\enddata
\tablenotetext{a}{Vacuum heliocentric.}
\tablenotetext{b}{$N$(C~IV) in cm$^{-2}$.}

\end{deluxetable}


\begin{deluxetable}{cllll}
\tighten
\tablecaption{\textsc{C~IV Absorption Systems in SDSS\,0836+0054}}
\tablehead{
  \multicolumn{1}{c}{Number}
& \multicolumn{1}{c}{$z_{\rm abs}$\tablenotemark{a}}
& \multicolumn{1}{c}{$b$}
& \multicolumn{1}{c}{$\log N{\rm (C~IV)}$\tablenotemark{b}}
& \multicolumn{1}{c}{~~Comments} \\
  \colhead{ }
& \colhead{ } 
& \multicolumn{1}{c}{(km~s$^{-1}$)}
& \colhead{ } 
& \colhead{ } 
}

\startdata
1       & $4.49951 \pm 0.00009$  & $15:$       & $12.50 \pm 0.08$  & \\ 
2       & $4.51440 \pm 0.00003$  & $41 \pm 2$  & $13.52 \pm 0.02$  & \\
3       & $4.61237:$             & \ldots      & $12.7:$           & ~~Marginal \\
4       & $4.66874 \pm 0.00005$  & \ldots      & $> 13.04$         & ~~Lines are saturated \\
5a      & $4.6826 \pm 0.0002$    & $71 \pm 15$ & $13.52 \pm 0.07$  & ~~Poor fit \\ 
5b      & $4.68487 \pm 0.00007$  & $39 \pm 8$  & $13.57 \pm 0.07$  & ~~Poor fit \\
5c      & $4.68690 \pm 0.00004$  & $35 \pm 4$  & $13.68 \pm 0.03$  & ~~Poor fit \\
6       & $4.7739 \pm 0.0001$    & $65 \pm 10$ & $13.15 \pm 0.05$  & ~~Marginal \\
7       & $4.99695 \pm 0.00005$  & $36 \pm 4$  & $13.56 \pm 0.03$  & \\
\enddata
\tablenotetext{a}{Vacuum heliocentric.}
\tablenotetext{b}{$N$(C~IV) in cm$^{-2}$.}

\end{deluxetable}


\begin{deluxetable}{cllll}
\tighten
\tablecaption{\textsc{C~IV Absorption System in SDSS\,1030$+$0524}}
\tablehead{
  \multicolumn{1}{c}{Number}
& \multicolumn{1}{c}{$z_{\rm abs}$\tablenotemark{a}}
& \multicolumn{1}{c}{$b$}
& \multicolumn{1}{c}{$\log N{\rm (C~IV)}$\tablenotemark{b}}
& \multicolumn{1}{c}{~~Comments} \\
  \colhead{ }
& \colhead{ } 
& \multicolumn{1}{c}{(km~s$^{-1}$)}
& \colhead{ } 
& \colhead{ } 
}

\startdata
1       & $4.94866 \pm 0.00006$  & $58 \pm 4$  & $13.98 \pm 0.02$  & \\ 
\enddata
\tablenotetext{a}{Vacuum heliocentric.}
\tablenotetext{b}{$N$(C~IV) in cm$^{-2}$.}

\end{deluxetable}


\begin{deluxetable}{llllcl}
\tighten
\tablecaption{\textsc{Statistics of C~IV Absorption}}
\tablehead{
  \multicolumn{1}{c}{Sample}
& \multicolumn{1}{c}{$z_{\rm abs}$}
& \multicolumn{1}{c}{$\Delta X$\tablenotemark{a}}
& \multicolumn{1}{c}{$\langle z_{\rm abs} \rangle$}
& \multicolumn{1}{c}{Number of} 
& \multicolumn{1}{c}{$\Omega_{\rm CIV}$\tablenotemark{b}}\\
  \colhead{ }
& \colhead{ } 
& \colhead{ }
& \colhead{ }
& \multicolumn{1}{c}{Lines}
& \multicolumn{1}{c}{($\times 10^{-8}$)}
}

\startdata
Whole sample    & 4.0--5.0 & 3.29  & 4.568  & 16 & $2.4 \pm 1.2$ \\
Subsample       & 4.5--5.0 & 1.87  & 4.688  & 11 & $3.6 \pm 2.1$ \\
Songaila (2001) & 4.5--5.0 & 5.36  & 4.655  & 16 & $2.5^{+1.9}_{-1.4}$ \\
\enddata
\tablenotetext{a}{For a flat cosmology with $\Omega_M=1$\,.}
\tablenotetext{b}{For a flat cosmology with $\Omega_M=1$ and
$H_0 = 65$\kms~Mpc$^{-1}$\,.
The error estimates on $\Omega_{\rm CIV}$ are 
90\% confidence limits.}
\end{deluxetable}


\begin{deluxetable}{ccc}
\tighten
\tablecaption{\textsc{Incompleteness Corrections}}
\tablehead{
  \multicolumn{1}{c}{$\log N$(C~IV)\tablenotemark{a}}
& \multicolumn{1}{c}{Fraction of C~IV Systems}
& \multicolumn{1}{c}{Correction Factor}\\
  \colhead{ }
& \multicolumn{1}{c}{Detected}
& \multicolumn{1}{c}{to $\Omega_{\rm CIV}$}
}

\startdata
14.0       & 1.0 & 1.0\\ 
13.7       & 1.0 & 1.0\\
13.5       & 1.0 & 1.0\\
13.3       & 0.96 & 1.01 \\
13.0       & 0.45 & 1.22\\
\enddata
\tablenotetext{a}{$N$(C~IV) in cm$^{-2}$.}
\end{deluxetable}

%
%

\clearpage
\begin{figure}
\hspace*{-3.5cm}
\psfig{figure=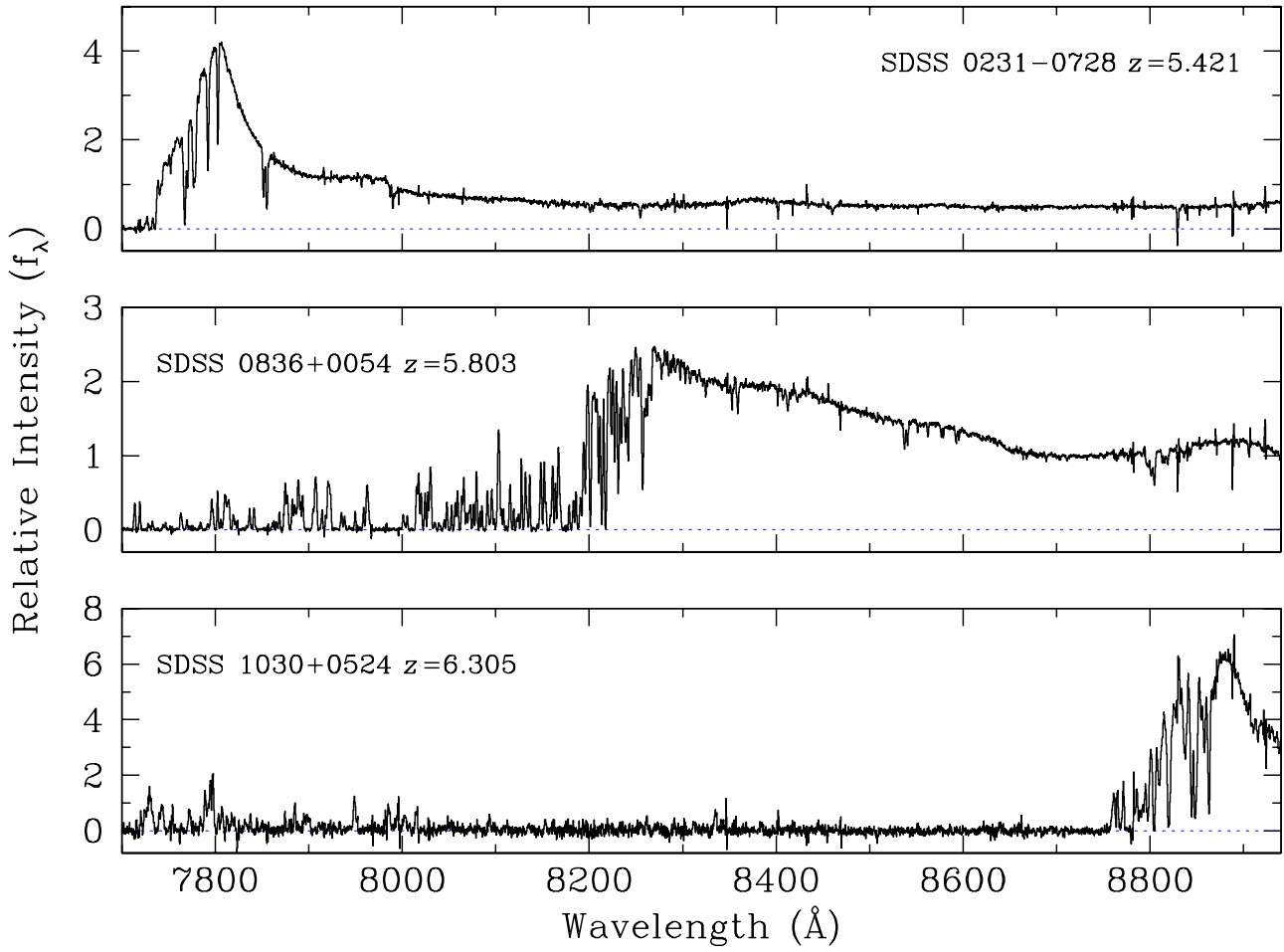,width=165mm,angle=270}
\figcaption{
Portions of the ESI spectra of the three QSOs
observed in this study.
} 
\end{figure}


\begin{figure}
\psfig{figure=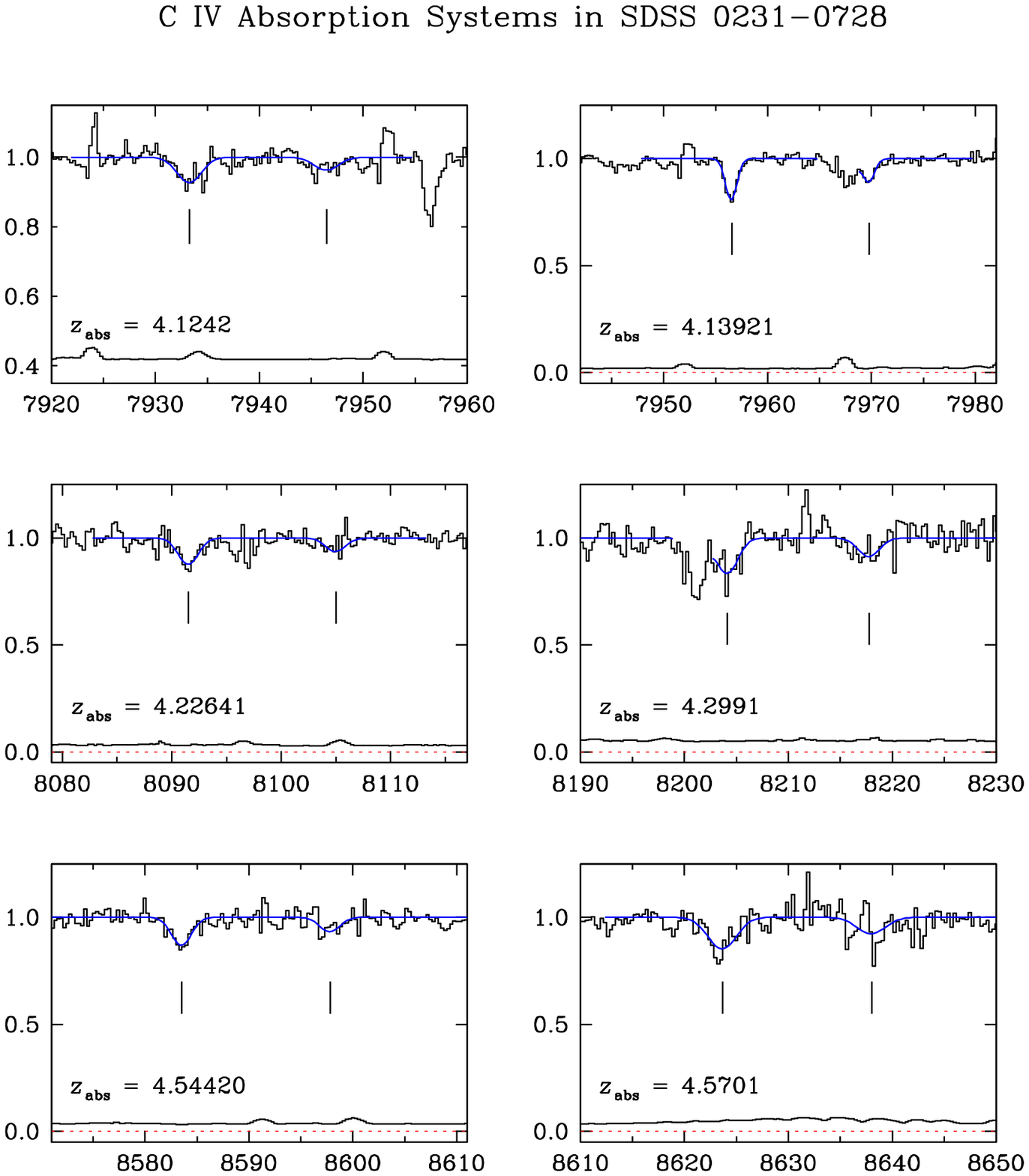,width=165mm}
\vskip -2.5cm
\figcaption{
Normalised absorption profiles (histograms)
of C~IV systems in SDSS\,0231$-$0728.
The $y$-axis is residual intensity, while the $x$-axis
gives the observed wavelength in \AA.
The thin continuous lines show theoretical profiles 
produced by VPFIT with the parameters listed
in Table 2. 
Vertical tick marks indicate the positions of 
the C~IV~$\lambda\lambda 1548.2041,1550.7812$ doublet lines.
The line near the zero level shows the $1 \sigma$ error
spectrum. The weak C~IV system at $z_{\rm abs} = 4.1242$
(top left-hand panel) is shown for clarity on an expanded
$y$-scale; in this case the error spectrum has been offset by 
$+0.4$ in $y$.
} 
\end{figure}


\begin{figure}
\psfig{figure=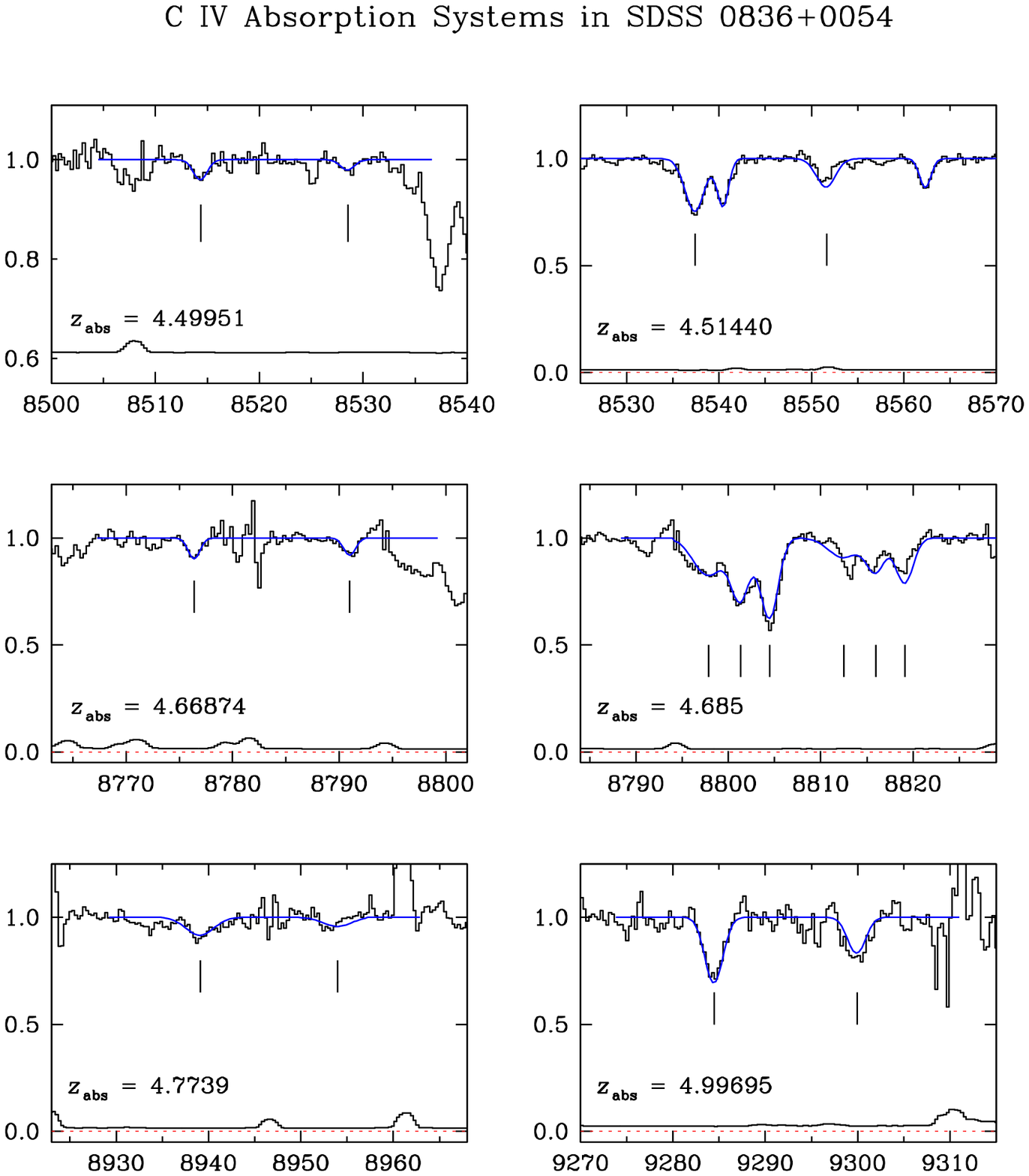,width=165mm}
\vskip -2.5cm
\figcaption{
Normalised absorption profiles (histograms)
of C~IV systems in SDSS\,0836+0054.
The $y$-axis is residual intensity, while the $x$-axis
gives the observed wavelength in \AA.
The thin continuous lines show theoretical profiles 
produced by VPFIT with the parameters listed
in Table 3. 
Vertical tick marks indicate the positions of 
the C~IV~$\lambda\lambda 1548.2041,1550.7812$ doublet lines.
The line near the zero level shows the $1 \sigma$ error
spectrum. The weak C~IV system at $z_{\rm abs} = 4.49951$
(top left-hand panel) is shown for clarity on an expanded
$y$-scale; in this case the error spectrum has been offset by 
$+0.6$ in $y$.
} 
\end{figure}


\begin{figure}
\psfig{figure=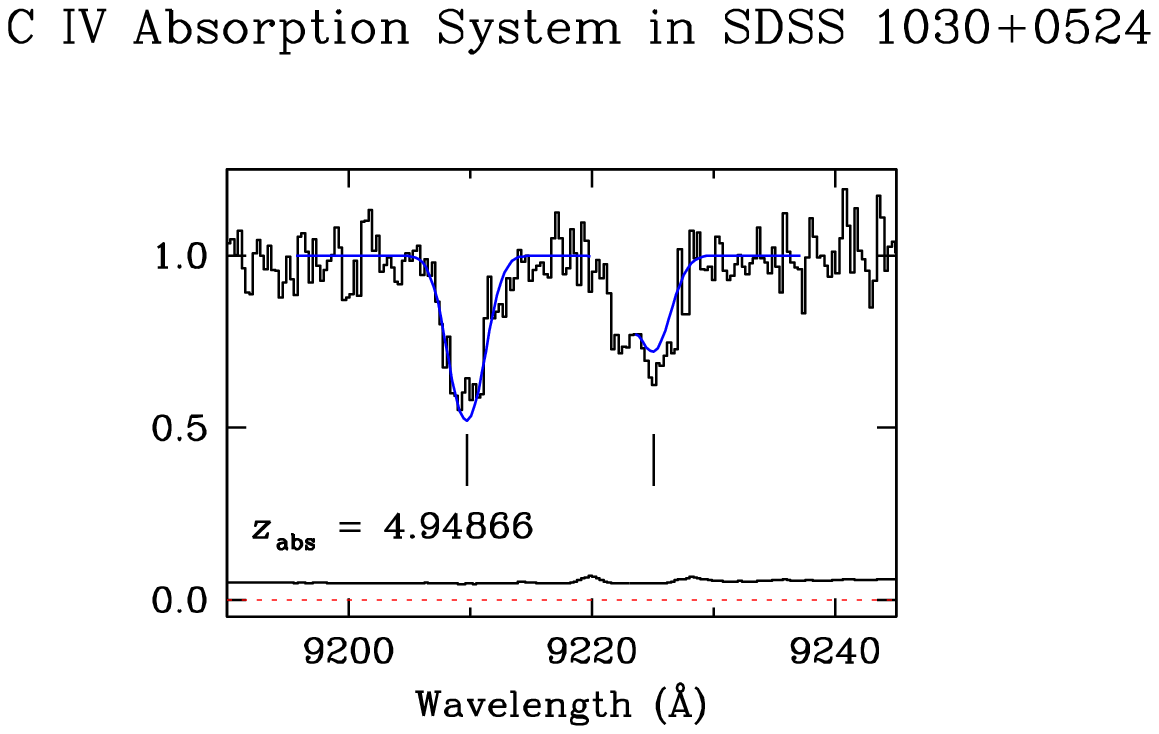,width=165mm}
\vskip -4.5cm
\figcaption{
Normalised absorption profile (histogram)
of the only C~IV system detected in SDSS\,1030+0524.
The $y$-axis is residual intensity.
The thin continuous line shows the theoretical profile
produced by VPFIT with the parameters listed
in Table 4. 
Vertical tick marks indicate the positions of 
the C~IV~$\lambda\lambda 1548.2041,1550.7812$ doublet lines.
The line near the zero level shows the $1 \sigma$ error
spectrum.
} 
\end{figure}


\begin{figure}
\vspace*{-2.5cm}
\hspace*{-0.4cm}
\psfig{figure=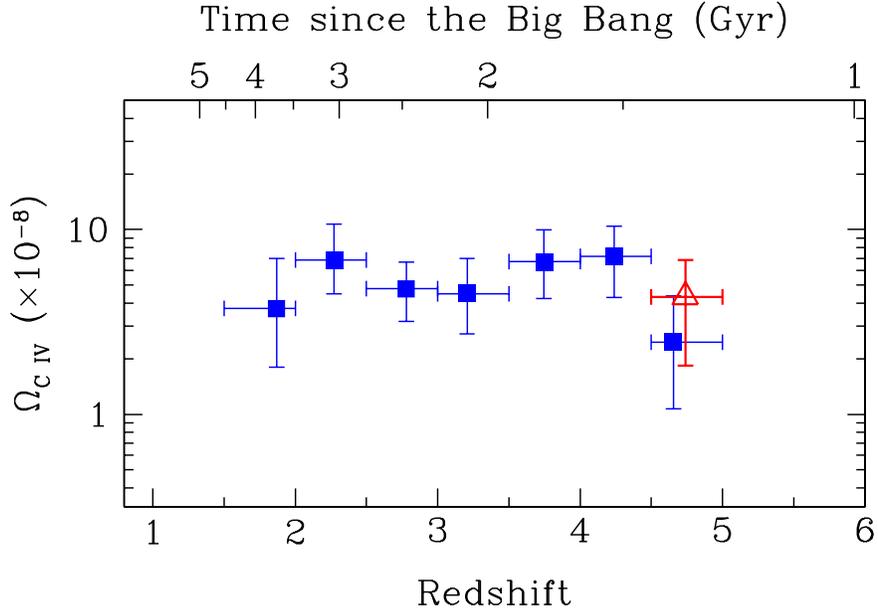,width=165mm}
\vskip -6.5cm
\figcaption{
Mass density of C~IV, $\Omega_{\rm CIV}$, plotted vs. redshift
and cosmic time. For consistency with earlier work,
$\Omega_{\rm CIV}$  was calculated from equation (1) and (3)
with $\Omega_{\rm M} = 1.0$ and $H_0 = 65$\,\kms~Mpc$^{-1}$.
(The time scale at the top of the figure, however, refers
to the currently favored 
$\Omega_{\rm M} = 0.3$ and $\Omega_{\Lambda} = 0.7$
cosmology).
Filled squares are the results by Songaila (2001),
while the triangle shows the single measurement 
from the present work.
Songaila estimates that at redshifts $z \leq 4$
her data are complete for column densities
$\log N$(C~IV)$ \geq 13$; for a proper comparison 
with these redshifts we have multiplied our own
measurement by a factor of 1.22 which takes into
account our incompleteness fraction above 
$\log N$(C~IV)$ = 13$, as discussed in \S5. For clarity,
the triangle has been shifted along the $x$-axis by 
+0.05 from its correct position at 
$\langle z \rangle = 4.688$ (see Table 5).
The error bars are 90\% confidence limits.
} 
\end{figure}

\end{document}